\documentclass[12pt]{article}

\usepackage{amsmath,amssymb}

\newcommand\al{\alpha}

\newcommand\de{\delta}

\newcommand\et{\eta}
\renewcommand\th{\theta}

\newcommand\rh{\rho}
\newcommand\si{\sigma}

\newcommand\vp{\varphi}

\newcommand\ps{\psi}
\newcommand\om{\omega}

\newcommand\De{\Delta}

\newcommand\Om{\Omega}

\newcommand\ie{\emph{i.e.}}

\newcommand\ea{\emph{et al}}

\newcommand\beq{\begin{equation}}
\newcommand\eeq{\end{equation}}
\newcommand\bal{\begin{align}}
\newcommand\eal{\end{align}}

\newcommand\rms[1]{_{\text{#1}}}

\newcommand\ap{\approx}

\newcommand\X{\times}
\newcommand\fr{\frac}

\newcommand\gap{\;\lower3pt\hbox{$\buildrel > \over
{\scriptstyle\sim}$}\;}
\newcommand\lap{\;\lower3pt\hbox{$\buildrel < \over 
{\scriptstyle\sim}$}\;}

\renewcommand\bal{\boldsymbol{\alpha}}

\newcommand\PL{{\em Phys.\ Lett.\ }}
\newcommand\PRD{{\em Phys.\ Rev.\ D\ }}
\newcommand\PRL{{\em Phys.\ Rev.\ Lett.\ }}

\newcommand\RPP{{\em Rep.\ Prog.\ Phys.\ }}

\newcommand\cM{\mathcal{M}}

\begin{document}

\rightline{Imperial/TP/041001}

\begin{center}
{\Large{\bf Cosmic Strings Reborn?\footnote{Invited Lecture
at COSLAB 2004, held at Ambleside, Cumbria, United Kingdom,
from 10 to 17 September 2004, with some later revisions}}\\
\vspace{12pt}
T.W.B.~Kibble}\\
{\large Blackett Laboratory, Imperial College\\
London SW7 2AZ, United Kingdom}\\
20 October 2004
\end{center}
\vspace{12pt}

\begin{abstract}
There are two main reasons for the recent renewal of
interest in cosmic strings:  Fundamental string-theory
models suggest their existence; and there are at least two
tentative observations of their possible effects.  In this
talk, I review their current status in the light of these
two factors.
\end{abstract}

\section{Introduction}

Cosmic strings were very popular in the eighties, and
much of the nineties \cite{HinK95,VilS94}, because they
seemed to offer a neat alternative to inflation as a means
of generating the primordial density perturbations from
which galaxies and clusters eventually grew
\cite{Bat+98,Ave+98}.  In particular, for GUT-scale
strings, the predicted string tension was about right to
explain their magnitude.  But towards the millennium their
popularity waned, swept away by the avalanche of data,
especially the microwave background measurements from COBE,
BOOMERanG, and, more recently, WMAP.  This eventually showed
beyond doubt that cosmic strings or other topological
defects could \emph{not} provide an adequate explanation for
the bulk of the density perturbations
\cite{Con+99,Bou+02,Dur+02}.  There always remained the
possibility that they might contribute at the level of a
few per cent, but that hardly seemed a sufficiently
exciting possibility to keep them alive.  There were only
27 papers published on cosmic strings in 2001, compared to
67 in 1997.

But against the odds, there has been a remarkable revival
of interest.  The number of papers on cosmic strings has
again begun to soar --- 46 in 2003.  There are two main
reasons, one theoretical, one observational.

When I gave talks about cosmic strings twenty years ago, I
always made a point of emphasizing that there was no
connection whatever between these hypothetical macroscopic
objects in the cosmos and the equally hypothetical
microscopic fundamental strings of superstring theory. 
The characteristic energy scales were very different, the
GUT scale or less for cosmic strings, something near the
Planck scale for fundamental strings.  But that is no longer
true.  The string scale may in fact be substantially
less \cite{Ark+98,Ark+99,RanS99}.  Moreover, string theory
or M-theory predicts, even demands, the existence of
macroscopic defects such as cosmic strings.  Branes, which
now play such a key role in string theory, are in essence
defects of various dimensions (not necessarily
topological).  In particular, in the brane-world picture,
colliding branes will in many cases generate cosmic strings
\cite{MajD02,SarT02,Pog+03,DvaV04}.  So the popularity of
string theory spills over to cosmic strings.  It is also
worth noting that supersymmetric GUTs, which may often form
a link between string theory and the standard model, seem to
demand cosmic strings \cite{Jea+03}.

On the observational side, there has been no unambiguous
detection of a cosmic string, but there have been
tantalizing hints --- observations whose most natural
interpretation seems to be in terms of cosmic strings. 
There are at least two of these.  Sazhin \emph{et al}
\cite{Saz+03,Saz+04} found a strange example, called CSL-1,
of a gravitational lens which seems to involve two images of
comparable magnitude of the same giant elliptical galaxy,
and have subsequently found an unexpectedly large number of
gravitational lens candidates in the vicinity.  Schild
\emph{et al} \cite{Sch+04} have found anomalous brightness
fluctuations in a multiple-image lens system, Q0957+561A,B,
which may be evidence for lensing by an oscillating loop of
cosmic string.

Here, I want to review the physics of cosmic strings, and
discuss their status in the light of these new
developments.

\section{Cosmic strings in the early universe}

Let me begin by reviewing a few basic facts about cosmic
strings in the early universe.

Cosmic strings are formed in many symmetry-breaking phase
transitions.  If the symmetry is broken from a group $G$
down to a subgroup $H$, the manifold of degenerate vacuum
or ground states is $\cM=G/H$, and the topology of this
manifold determines the types of defect that can form
\cite{Kib00}.  In particular, strings can form if $\cM$ is
not simply connected, \ie, if its first homotopy group
$\pi_1(\cM)$ is non-trivial.

There are many variants on the theme of cosmic strings --
composite defects, embedded defects, semilocal strings,
superconducting strings, and so on.  I shall not have much
to say about any of these, though no  doubt others will.

From a cosmological point of view, most attention has been
paid to GUT strings.  In the standard model of particle
physics, based on the symmetry group $G_{321}=$
SU(3)$\X$SU(2)$\X$U(1), the energy dependence of the
coupling strengths $g_3,g_2,g_1$ of the three interactions
at laboratory energies suggests that all three may be
united in a \emph{grand unified theory} (GUT) at an energy
scale of $10^{15}$ or $10^{16}$ GeV, at least provided we
include supersymmetry.  GUTs have been proposed based on
simple groups $G$, such as SU(5) or SO(10), or on
semisimple groups with extra discrete symmetries, like the
left--right symmetric $G=$ SU(4)$\X$SU(4).  In unified
theories derived from string theory even larger groups
appear.

Symmetry breaking can occur in one or more stages from $G$
down to $G_{321}$.  In many cases, this symmetry breaking is
associated with topological defect formation.  Almost
always, monopoles are formed at one of the transitions, and
must be eliminated in some way, for example by an
intervening period of inflation.  But more to the point for
the present discussion, cosmic strings are often formed.

One of the things that initially made the idea of cosmic
strings so exciting was an apparent coincidence of scales. 
They seemed to offer an explanation for the magnitude of the
initial density perturbations from which galaxy and
clusters eventually grew.

The symmetry-breaking scale $\et$ determines the critical
temperature $T\rms{c}$ for the transition.  Also, those
components of the gauge field that correspond to broken
symmetries acquire masses $m_X$ of this order (multiplied
by appropriate coupling constants).  In these relativistic
strings, because of Lorentz invariance under boosts along
the direction of the string, the mass per unit length and
the string tension are equal.  This tension
$\mu$ is of order
 \begin{equation}
\mu\sim\et^2\sim m_X^2.
 \end{equation}

The most important observational consequences of cosmic
strings (so long as they are non-superconducting) stem from
their gravitational effects, whose strength is characterized
by the dimensionless constant
 \begin{equation}
G\mu\sim\fr{\et^2}{m\rms{Pl}^2},
 \end{equation}
where $G=1/m\rms{Pl}^2$ is Newton's constant.  In
particular, as I'll discuss shortly, cosmic strings
generate density perturbations in the universe with a
typical amplitude $\de\rh/\rh\sim G\mu$.

The key point is that for $\et$ or $m_X$ of the order of
the GUT scale, we have
 \begin{equation}
G\mu\rms{GUT}\sim10^{-6}\text{ to }10^{-7},
 \end{equation}
so the density perturbations predicted are at least in the
right ballpark to provide a resolution of one of the
long-standing puzzles of cosmology: where do the
primordial density perturbations originate, from which
galaxies and clusters eventually evolve?

Through the 80s and into the 90s, a huge amount of work was
done to try to put flesh onto the bones on this appealing
idea.  A clear picture of cosmic string evolution gradually
emerged, the result of both analytic studies and numerical
simulations by several different groups.  (For references,
see \cite{HinK95,VilS94,Aus+95}.)  We now think we have a
pretty good idea of what will happen to a network of cosmic
strings formed at an early phase transition, and how it
might influence the cosmology --- though it must be said
there is still room for doubt about some of the details. 

We start with a random tangle of strings, characterized by a
length scale $\xi$, which may be defined as the length such
that in a typical volume $\xi^3$ we expect to find a length
$\xi$ of string.  In other words, the strings contribute a
mean density
 \begin{equation}
\rh\rms{str}=\fr{\mu}{\xi^2}.
 \end{equation}
Equivalently, $\xi$ is roughly the typical distance between
strings.  The initial value of $\xi$ is determined by the
interplay between microphysics and cosmic expansion. 
Typically we expect $\xi$ to be much smaller than the
Hubble radius at the time.

The key to the subsequent evolution is the
\emph{topological stability} of the strings.  They may
stretch or shrink, but they cannot break, though when they
meet they can exchange partners or \emph{intercommute}.  As
the universe expands, the strings are stretched, and kinks
are gradually straightened.  But when the strings
intercommute, new kinks are formed.  If there were no other
energy-loss mechanism, the strings would eventually come to
dominate the energy density of the universe.  But there is
an energy-loss mechanism --- the formation and decay of
small loops.  When a string intersects itself, it cuts off
a closed loop.  Once formed, a loop is doomed unless it
happens to reconnect with a longer piece of string (which
for small loops is improbable).  The loop oscillates and
gradually loses energy by gravitational radiation, until it
disappears altogether.  (A different view of the evolution
is presented in another contribution to this meeting
\cite{Hin04}.)

In this process, the characteristic length scale $\xi$ of
the network grows faster than the Hubble radius.  It is
believed to lead eventually to a scaling regime, in which
$\xi$ grows proportionately to the Hubble radius $1/H$ or
the age of the universe, $t$:
 \begin{equation}
\xi\propto1/H\sim t.
 \end{equation}
Consequently, in this regime the strings contribute a
constant fraction of the critical density:
 \begin{equation}
\Om\rms{str}=\fr{8\pi G}{3H^2}\rh\rms{str}\sim G\mu.
 \end{equation}
However, the numerical coefficient here does change
significantly when we pass from a radiation-dominated to a
matter-dominated universe.

One important side effect of the breaking off of small
loops is that the strings become quite kinky on small
scales.  Typically, the distance between kinks is much
smaller than $\xi$.  This also means there is substantial
loss of energy by gravitational radiation from the long
strings as well as the loops.

\section{Cosmological effects}

The next stage, once we had a fair idea of how strings
evolve, was to estimate their likely cosmological effects. 
This inevitably involved a lot of numerical work, taking as
input the results of simulations of the string evolution
process, and using them to predict the large-scale
density perturbations explored by galaxy distribution
surveys, and the temperature inhomogeneities in the cosmic
microwave background (CMB).

It gradually became apparent, however, that although the
predictions were in the right ballpark, it was very
difficult to get them to fit precisely, especially to
fit both the large scale structure and the CMB
simultaneously \cite{Con+99,Bou+02,Dur+02}.

Meanwhile, the rival inflationary theory --- the idea that
the origin of the primordial density perturbations can be
traced back to quantum fluctuations during a very early
period of inflation --- was having much more success, in
particular in fitting the peaks of the angular power
spectrum, as measured by COBE and other instruments.  These
peaks can be traced back to the fact that all the Fourier
modes of the density perturbation started out essentially
in phase at the end of inflation.  Their positions and
heights are very characteristic of the inflationary
scenario.

The \emph{coup de gr\^ace} was provided by WMAP in 2003,
with the first measurements of the CMB polarization, and the
temperature--polarization cross-corre\-lation, which fitted
the inflationary predictions excellently, and could not be
well explained by cosmic strings or other defects
\cite{Pei+03,Bev+04}.

Of course, all this doesn't prove there are no cosmic
strings.  It only proves that if they do exist they don't
contribute more than an insignificant proportion of the
primordial density perturbation.  The observations give an
upper limit on the value of the parameter $G\mu$.  Pogosian
\emph{et al} \cite{Pog+03} quote a limit
 \begin{equation}
G\mu\le 1.3\X10^{-6}\sqrt{\frac{B\lambda}{0.1}},
 \end{equation}
where $\lambda$ is the probability of intercommuting when
two strings meet (most often assumed to be 1), and
$B$ is the fraction of the CMB power spectrum
attributable to cosmic strings, which certainly satisfies
$B<0.1$.  On the other hand, a recent study by Jeong and
Smoot \cite{JeoS04}, searching for evidence of a
cosmic-string contribution in the WMAP data, has yielded the
tighter bound
 \begin{equation}
G\mu\le 3.3\X10^{-7}.
 \end{equation}
If this is correct, as we shall see, it is at least
marginally in conflict with the reported observational
evidence for detection of cosmic strings; it is, however,
somewhat dependent on assumptions about string evolution.

Another, possibly even stronger limit on $G\mu$ can be
obtained from limits on the gravitational radiation that
would be emitted by cosmic strings.  Observations of the
regularity of pulsar timing put an upper limit on the
fraction of the critical density in gravitational waves
with periods of up to 10 years \cite{Lom02}, 
 \begin{equation}
G_{\textrm{gw}}\lap4\X10^{-9}.
 \end{equation}
An estimate of the gravitational radiation emitted by
strings (see for example Ref.~\cite{VilS94}) suggests that
this implies a bound
 \begin{equation}
G\mu\lap10^{-7}.
\label{gwlim}
 \end{equation}
If correct, this is in clear contradiction with the
interpretation of observational evidence discussed later
as due to cosmic strings.  However, one should be cautious;
the estimate is subject to considerable uncertainties. 
Because of the huge range of scales involved, one of the
least certain aspects of cosmic-string dynamics is the
long-term evolution of the small-scale structure on
strings.

For obvious reasons, the discovery that cosmic strings could
not provide an explanation for the primordial density
inhomogeneities led to a widespread loss of interest in the
whole idea.  Why bother with something for which no real
evidence exists, when a different theory provides a very
satisfying and astonishingly accurate description of the
data?

So why now the revival of interest?

\section{Cosmic strings in string theory}

As I said at the start, there are two main reasons.  The
first, and probably the most important, is that string
theory cosmologists have discovered cosmic strings lurking
everywhere in the undergrowth.  These connections between
fundamental strings and cosmic string will be discussed in
this session by several other people, much better qualified
to do so than myself.  So I shall confine myself to a few
remarks.  (For a recent review, see Ref.~\cite{Pol04}.)

The big problem with string theory has been to connect its
beautiful mathematical structure to any real measurements
or observations.  The chain of reasoning from fundamental
string theory to phenomenology or cosmology is long, and it
has proved hard to find definitive observational tests, or
even to suggest where one might look.  But it now appears
that cosmic strings might actually provide the best
observational window into fundamental string theory.  There
are several reasons for this.

Firstly, there is the question of string scale.  The string
tension of fundamental strings is the square of the string
energy scale, which always used to be identified with the
Planck scale.  Such heavy strings certainly do not exist in
our universe today, and cannot have played any role in
cosmological evolution except conceivably in the first few
Planck times.  But we now know of models with large compact
dimensions, in which the string scale may be much lower,
down to the GUT scale or even less
\cite{Ark+98,Ark+99,RanS99}.  So in principle fundamental
strings of macroscopic length are not ruled out.

Secondly, string theory now provides a much richer family
of defects \cite{DvaV04,Cop+04} --- not only fundamental
(F-) strings but also D-branes of all dimensions.  These can
include D-strings whose ends are tied to D-branes of higher
dimension.  And there may also be D-branes partly wrapped
on the compact dimensions, which look like strings on a
macroscopic scale \cite{DvaV04,Hal04}.  Some of the
resulting cosmic strings have novel properties --- for
example, a probability less than unity of intercommuting
when they meet \cite{Jon+03,Jac+04}.  One can also have
defects that are composites of F- and D-strings, which can
form complex networks with vertices where three strings
meet.  Which of these possibilities is actually realized
depends on the precise form of the dimensional reduction
from 10 or 11 dimensions to 4.  For that reason, if we
could discover what kinds of macroscopic defects exist in
our universe it would tell us a lot about the underlying
fundamental theory.

Finally, it now appears that GUTs will almost inevitably
lead to cosmic strings.  In the long road from string
theory or M-theory to the standard model, it is very
natural to go through an intermediate GUT stage. 
Successful grand unification seems to demand supersymmetry
--- without it the running couplings of the three
fundamental interactions do not converge at a single
energy.  Since grand unification does inevitably produce
monopoles, it also demands an inflationary period to
eliminate them, and of course inflation is needed for
other reasons too.  In a very interesting recent study,
Jeannerot \ea\ \cite{Jea+03} looked exhaustively at all
possible simple GUT groups of rank $\le8$, and at all
possible symmetry-breaking chains down to the $G_{321}$ of
the standard model.  Many of these can be ruled out because
their predictions are in conflict with observation, for
example by predicting the formation of monopoles or domain
walls \emph{after} inflation.  The remarkable fact is that,
although many possibilities remain, every one of them
predicts the formation of topological or embedded cosmic
strings at the end of inflation.  So it seems that cosmic
strings are almost unavoidable.

\section{Gravitational effects of cosmic strings}

Now let me turn to the other main reason for the renewed
interest in cosmic strings: the tantalizing hints that
their effects may already have been detected.  All these
possible observations of cosmic strings depend on their
very distinctive gravitational lensing effects.

The space-time around a straight cosmic string is very
unusual.  Because of the equality between energy per unit
length and string tension, nearby masses experience no
gravitational acceleration towards the string.  The space
is locally flat, but globally curved.  In fact, it is
cone-shaped, with curvature confined to the core of the
cosmic string.  For a string along the $z$ axis in
otherwise empty space, the surrounding space-time metric is
 \begin{equation}
ds^2=dt^2-dz^2-d\rh^2-(1-8G\mu)\rh^2d\vp^2,
 \end{equation}
(assuming $G\mu\ll1$).  Equivalently, the local flatness
can be made explicit by introducing a new angle coordinate
$\bar\vp=(1-4G\mu)\vp$, so that the metric becomes 
 \begin{equation}
ds^2=dt^2-dz^2-d\rh^2-\rh^2d\bar\vp^2.
 \end{equation}
But here $\bar\vp$ runs not from $0$ to $2\pi$ from $0$ to
$(2\pi-\de)$, where the defect angle $\de$ is
 \begin{equation}
\de=8\pi G\mu\ap5''\!\!.2\left(\fr{G\mu}{10^{-6}}\right).
 \end{equation}
For a GUT scale string, this angle is a few seconds of arc.

This means that a straight string acts like a
cylindrical gravitational lens with a very unusual and
characteristic pattern of lensed images.  In general, we
should see two images of a source behind the string,
separated by an angle of order $\de$.  More precisely, the
separation of the images is \cite{Vil91,Got95}
 \begin{equation}
\al=\fr{D_{ls}}{D_s}\de\sin\th,
\label{imagesep}
 \end{equation}
where $D_s$ is the angular diameter distance of the source
from us, $D_{ls}$ is that of the source from the lens, and
$\th$ is the angle between the line of sight and the
tangent to the string.  In sharp contrast to the lensing
pattern of ordinary gravitational lenses, the string
induces no magnification or demagnification, and the two
images are of equal magnitude (unless one of them is only a
partial image).

In practice, the situation will be more complicated, for
at least two reasons.  First, the strings are in general
moving, often with quite large velocities.  What is
significant is the velocity component perpendicular to the
line of sight.  Suppose the string is moving with transverse
velocity $v$, or equivalently, in the rest frame of the
string, both observer and source are moving with velocity
$-v$.  Then the two images will have slightly different
red-shifts: the image behind the string will be
blue-shifted relative to that ahead of it, leaving a
frequency difference $\de\om$ of order
 \begin{equation}
\fr{\de\om}{\om}\sim v\de,
 \end{equation}
and the formula for the separation angle (\ref{imagesep})
is more complicated.  This also applies to the CMB: a
transversely moving string would induce a discontinuity in
the temperature of the CMB \cite{KaiS84}, a unique signal of
a cosmic string if it could be observed.

Secondly, as I explained earlier, the strings are not
straight but kinky.  This means that, viewed on a coarser
scale, the effective energy per unit length $U$ and string
tension $T$ are no longer equal \cite{VacV91}.  For example,
if the string is a zigzag of straight segments each making
an angle $\ps$ with a median line, then
 \begin{equation}
U=\mu\sec\ps,\qquad T=\mu\cos\ps.
 \end{equation}
Note that
 \begin{equation}
U>\mu>T,\qquad\text{and}\qquad UT=\mu^2.
 \end{equation}
These are in fact general results \cite{Vil90,Car90}.

Because of this, the ordinary gravitational attraction
towards the string no longer vanishes; instead it is
 \begin{equation}
g=\fr{2G(U-T)}{r},
 \end{equation}
where $r$ is the distance from the string.  Moreover, the
defect angle is now
 \begin{equation}
\de=4\pi G(U+T).
 \end{equation}
However, because of the gravitational attraction, the
actual separation of the images is larger than this would
suggest, namely
 \begin{equation}
\al=\fr{D_{ls}}{D_s}8\pi GU\sin\th.
\label{imagesep2}
 \end{equation}

\section{Possible observation of a cosmic string}

Some years ago, possible observations of lensing by cosmic
strings were reported \cite{CowH87,Hu90}, but did not
apparently stand up to further scrutiny.  Recently,
however, there have been two new observations that seem to
suggest the presence of strings.  I would like to devote the
remaining part of this talk to an examination of these
claims.

The first is the result of an Italian--Russian
collaboration, in which Sazhin \ea\ \cite{Saz+03} report the
observation of a lensing candidate named CSL-1
(Capodimonte-Sternberg Lens Candidate no.~1).  There are
three other candidates, named CSL-2 to CSL-4, that have not
yet been fully analyzed. This candidate consists of a pair
of galaxy images found in the OACDF survey
(\emph{Osservatorio Astronomico di Capodimonte --- Deep
Field}).  The two images, separated by
$2''$, or approximately 20 kpc, look nearly identical, both
have a red-shift of $z=0.46\pm0.008$, and their magnitudes
in three different frequency bands are equal within
errors: 

 $$\begin{array}{llll}
&\qquad\text{B}&\qquad\text{V}&\qquad\text{R}\\
m_A&22.73\pm0.15&20.95\pm0.13&19.67\pm0.20\\
m_B&22.57\pm0.15&21.05\pm0.13&19.66\pm0.20
 \end{array}$$
\vspace{6pt}

One hypothetical explanation might be that this is an image
of a single large galaxy, with the central part of the image
obscured by a dust lane.  It would of course be a
remarkable coincidence for this to leave two equal images,
and in any case the authors rule out this possibility by
examining the spectral profile.

Another possibility which cannot be ruled out is that what
we see are two almost identical galaxies that just happen
to be one nearly behind the other and quite close together
in red-shift space.  However, this would require a
remarkable coincidence.

If the images are indeed of the same galaxy, it is
theoretically possible that this is due to some more
conventional lensing object, but in that case two identical
images would be extremely unlikely.  Moreover, the authors
show that it would have to be a giant galaxy, which should
be readily visible, and no such object is seen.  So they
conclude that the most likely explanation is lensing by a
cosmic string.  If so, we should expect
 \begin{equation}
G\mu\ge 4\X10^{-7},
 \end{equation}
which appears at least marginally in conflict with the limit
set by Jeong and Smoot \cite{JeoS04}, and even more so
with the limit (\ref{gwlim}) derived from gravitational
radiation.  Moreover, $G\mu$ would have to be \emph{larger}
than this if either of the other factors in
(\ref{imagesep2}), $D_{ls}/D_l$ or $\cos\th$, is
substantially less than unity.  On the other hand, it
should also be noted that it is really a limit on
$GU$; if the strings are kinky, $G\mu$ could be less.

Sazhin \emph{et al} \cite{Saz+03} pointed out that the
hypothesis that this is an example of lensing by a cosmic
string should in principle be easy to test by doing more
precise photometric studies.  They also point out that if
there is a string between us and the source galaxy it should
be lensing the CMB, producing a line of discontinuity, so
high-resolution radio-frequency measurements would also be
useful.

Meanwhile, they have produced additional evidence of
different kind \cite{Saz+04}.  If this is a genuine case of
lensing by a cosmic string, there should be other lensed
pairs in the vicinity \cite{HutV03}.  So they looked at
images of galaxies in a $4000\X4000$-pixel section of the
field ($16'\X16'$) centred on CSL-1 to see how many possible
examples they could find.

They estimate, based on other surveys, that in this region
there should be approximately 2200 galaxies within the
magnitude range 20 to 24 (in the R band).  This is indeed
roughly what they do see.  They then ask how many of these
should be lensed by a cosmic string across the field. 
Roughly speaking, all galaxies within a strip of width
$2\de$ centred on the string should produce double images. 
How long the strip is depends of course on how straight the
string is on the relevant scale, but the extreme limits
are a straight string and a random walk \cite{HutV03}.  So
they estimate that the number of lensed pairs of galaxies
should lie between 9 (for a straight string) and around 200
(for a random walk).  By contrast, in the same area they
expect no more than two lensed pairs due to conventional
lensing objects such as galaxies.

They then looked at pairs of objects with angular
separations between $1''$ and $4''\!\!.5$, and used a
statistical test due to Schneider \emph{et al}
\cite{Sch+92}, based on matching colours, to decide whether
these pairs were in fact images of the same object.  They
found 11 classified as very likely candidates, several
times what would be expected from conventional lenses. 
This provides added weight to the hypothesis that a cosmic
string has been seen.  The authors emphasize, however, that
to confirm that these really are lensed pairs,
spectroscopic analysis will be required.

Another important piece of information concerns the
distribution of the possible image pairs across the
$16'\X16'$ section of the field.  Given that the number is
not much larger than would be expected for a straight
string, we might expect that they are concentrated on a
roughly linear strip.  That does not seem to be the case
\cite{Saz04,Saz+04b}, but equally they do not appear to be
scattered randomly.  They could perhaps match a string with
a couple of kinks, zigzagging across the field.  Another
useful test would be to look for similar image pairs in a
quite different region of the field where there is no
lensing candidate.

To be fair, we should also note that there have been other
similar searches for lensed pairs that have not yielded
positive results, for example that of Shirasaki \ea\ 
\cite{Shi+03}.  But of course, that was in four randomly
chosen patches of the sky, not one already including a
cosmic-string candidate.

\section{Possible observation of a loop}

The other intriguing piece of evidence comes from an
analysis by Schild \ea\ \cite{Sch+04} of brightness
fluctuations in a very well known gravitational lens
system, Q0957+561, that has been studied intensively for 25
years.  It consists of two quasar images separated by
approximately $6''$.  They are known to be images of the
same quasar not only because of the spectroscopic match,
but also because the images fluctuate in brightness, and
the time delay between fluctuations is always the same; a
brightening of $A$ is matched by a brightening of $B$ 417.1
days later.  This pair is famous because the time delay
has been used to provide a measure of the Hubble constant,
independent of the distance ladder based on Cepheid
variables.  The reason for the lensing is well known ---
it is a foreground galaxy readily visible between the two
images, about $1''$ from $B$.

The brightness curves do not of course match precisely; the
images also fluctuate independently, principally due to
microlensing by individual stars in the lensing galaxy. 
However, what Schild and colleagues have found is evidence
of another component in the fluctuation, in which the
images fluctuate synchronously, with no time delay.  A plot
of the simultaneous magnitude fluctuations for the period
between September 1994 and July 1995 seems to show a
short-lived, quasi--periodic, synchronous fluctuation, with
an amplitude of about 0.2 magnitudes and a period that
appears from the figure to be about 80 days (though the
authors quote 100 days), lasting for three or four periods.

For comparison, they show the data for $A$ for the same
period compared with those for $B$ 417 days later, and also
the data for $B$ matched against those for $A$ 417 days
earlier.  Neither of these plots shows the degree of
correlation found in the first figure, although, as might
be expected, both do (at least to me) seem to show
\emph{some} correlation.  It is difficult to judge by
visual comparison how strong the evidence for synchronous,
quasi-periodic oscillations really is.  Certainly it is not
as yet conclusive.  As the authors point out, we lack a
proper statistical test for the hypothesis.

However, let us assume that the oscillation is real.  It
would of course be very difficult to explain on the basis
of microlensing, because it would require an accidental
coincidence of timing for at least three successive pairs of
microlensing events.  So what is the alternative?  The
authors suggest that the effect may be due to lensing by an
oscillating loop of cosmic string between us and the
lensing system. 

To get some estimate of the effect, Schild \emph{et al}
rely on earlier work of De Laix and Vachaspati
\cite{deLV96}.  They look at a particularly simple model, a
loop comprising a rotating double line, rotating in the
plane normal to the line of sight about its mid-point.  The
position of the loop is given by
 \begin{equation}
 \begin{array}{rl}
x\!\!\!&=R\cos\dfrac{t}{R}\sin\si,\\[8pt]
y\!\!\!&=R\sin\dfrac{t}{R}\sin\si,
 \end{array}
 \end{equation}
where the parameter $\si$ runs from $0$ to $2\pi$. 
Obviously, the oscillation period (half the rotation period)
is $\pi R$.  If this is to match the observed period of the
brightness fluctuations, 100 days, one finds
 \begin{equation}
R\ap0.02\text{ pc}\ap4400\textsc{ au}.
 \end{equation}
To have smooth quasi-periodic oscillations, $R$ should be
subtend an angle $\th_R$ substantially smaller than the
angle between the images (otherwise there would be sharp
spikes or discontinuities in the record), so this suggests
$\th_R\sim1.5''$, whence the distance of the string from us
is 
 \begin{equation}
D\rms{loop}\sim3\text{ kpc}.
 \end{equation}
This is remarkably close; if there is a loop this near us,
well inside our galaxy, either the density of loops is much
higher than generally expected, or we have been very
fortunate!

The fact that we saw only a few oscillations suggests that
the loop is moving across our field of view, possibly with
a velocity $v$ of order $0.7c$.

The expected size of the magnitude fluctuation for this
simple case is 
 \begin{equation}
\De m\ap\fr{384\pi^2(G\mu)^2\th_R^4}{\th_I^6},
\label{loopmag}
 \end{equation}
where $\th_I$ is the angular impact distance of the line
of sight relative to the centre of the loop.  The authors
suggest that $\th_I$ should be about half the distance
between the images, say $\th_I\ap3''$.  This allows an
estimate $G\mu\ap3\X10^{-8}$.  However, the very strong
dependence of (\ref{loopmag}) on the poorly known
parameters $\th_R$ and $\th_I$ means that this can only be
a very rough estimate.  The authors say that a more
accurate calculation based on simulations leads to
$G\mu\ap6\X10^{-7}$, but there must still be a large error
on this figure.

There is at first sight a possible alternative to the
hypothesis of lensing by an oscillating loop, namely
lensing by a double star.  However to get the required
period and amplitude, the stars would have to be at least
1.2 pc from us, with an orbital radius of $\sim1.8$ {\sc
au}.  That would mean masses of order $78M_\odot$.  It is
inconceivable that such a large pair of stars could have
escaped detection.  So if the synchronous magnitude
fluctuations are real, a cosmic string loop very near us
may be the least implausible explanation.

\section{Conclusions}

Modern string theory certainly encourages the belief that
cosmic strings should make an appearance in the early
universe, even if they are not the primary source of
initial density perturbations.  There is very good reason
for continuing to search for them.

As to whether they have already been found, the jury is
still out.  Speaking personally, it seems to me that a
reasonably strong case has been made for the hypothesis that
the lensing candidate CSL-1 is an example of lensing by an
intervening string, though serious questions remain to be
answered.  In any case, more detailed studies of the other
possible lensed pairs in the vicinity should put the
question beyond doubt.  So far as the synchronous
oscillations observed in the lensed quasar pair are
concerned, the explanation in terms of a very nearby
oscillating loop seems to require rather a lot of special
circumstances.  The real question is: are these apparent
synchronous oscillations a real effect, or the result of
chance fluctuations?  We certainly need a good statistical
test.

Whether these pieces of evidence stand up or not, the
search for cosmic strings will certainly continue for some
time to come.  If it turns out that these observations are
not due to cosmic strings, and if the upper limit on $G\mu$
drops significantly below $10^{-7}$, then direct detection
via gravitational lensing is unlikely to be feasible.  In
that case, the most likely way of verifying their existence
would be via the gravitational radiation they emit; they
should be detectable by LIGO or LISA \cite{Pol04}.  But
more theoretical work is needed to tie down the
predictions. 

\section*{Acknowledgements}

This work was supported by the European Science Foundation
through the \emph{Cosmology in the Laboratory} (COSLAB)
Programme.  I am indebted to the organizers of the meeting
for inviting me to present this review, and to several
participants, especially T.~Vachapati, as well as to
J.~Polchinski and M.~Sazhin, for helpful comments.

\end{document}